\documentclass[letterpaper,11pt]{article}
\usepackage[margin=1in]{geometry}
\usepackage{setspace,graphicx,url,hyperref,natbib,mathtools}
\usepackage{algorithm,algpseudocode}
\usepackage{amsmath,amsthm,amssymb,amsfonts}
\newcommand{\N}{\ensuremath{\mathbb{N}}}
\newcommand{\R}{\ensuremath{\mathbb{R}}}

\begin{document}
\title{Scalable computation of the maximum flow in large brain connectivity networks}
\author{Jingyun Qian and Georg Hahn}
\date{}
\maketitle

\doublespacing
\begin{abstract}
Concepts borrowed from graph theory, such as the maximum flow, have been used to study quantities such as information flow and information routing in brain connectivity networks. However, the runtime of $O(|V||E|^2)$ for the classic Edmonds-Karp algorithm renders computations of the maximum flow in networks $N=(V,E)$ with thousands or millions of vertices infeasible, where $V$ is the vertex set and $E$ is the edge set. We are interested in approximating the maximum flow in large connectivity networks. In this work, we propose a simple Monte Carlo algorithm to approximate the maximum flow in large networks via subsampling. Our algorithm returns a point estimate of the maximum flow with confidence bounds, and its runtime scales independently of $V$ or $E$ as $O(B \cdot |\tilde{V}| |\tilde{E}|^2)$, where $B$ is the number of Monte Carlo samples, $\tilde{V}$ is the set of subsampled vertices, and $\tilde{E}$ is the edge set induced by $\tilde{V}$. We evaluate our proposed algorithm with respect to both accuracy and runtime on simulated networks as well as networks downloaded from the Brain Networks Data Repository (\url{https://networkrepository.com/}).
\end{abstract}

\section{Introduction}
\label{sec:introduction}
Recent years have seen the advent of an exciting area of research focused on understanding the human brain \citep{Liang2024,Mosharov2025,Rittmo2026}. Apart from medical and biological aspects, the natural network structure induced by the neurons and synapses in the brain inspired new graph theoretical approaches to model the brain and investigate some of its properties. Previous work has investigated the relationship of information transfer to functional connectivity \citep{Ito2017,Neudorf2023}, the length of paths of cortical regions in the brain \citep{Rodrigues2009,Budd2012,Jahanshad2012}, the routing of communication in a network \citep{Seguin2018,AK2019,LiangMa2024}, or resting state functional connectivity \citep{Wu2021,Goni2023}.

This article focuses on information flow and information routing, which can be assessed in a suitably defined network via a computation of the maximum flow. In a capacitated and directed network, the maximum flow from a source node to a sink node is defined as the maximum amount that can be pushed from the source to the sink via any directed path. The interpretation of the capacities is context dependent. In our application to brain connectivity networks, the capacities usually encode some measure of information content that can be routed from one part of the brain to another. The most prominent algorithms for computing the maximum flow in a capacitated and directed network are the ones of \cite{Ford1956} and \cite{Edmonds1972}.

These classic algorithms for computing the maximum flow are not suitable for large scale brain connectivity networks with thousands or millions of vertices due to their unfavorable runtime scaling. In particular, the runtime of the classic Edmonds-Karp algorithm is $O(|V||E|^2)$, where $V$ is the set of vertices and $E$ is the set of edges. Under the assumption of $|E| \simeq |V|^2$ for dense networks, this simplifies to a quintic runtime in $|V|$.

In this contribution, we propose a new Monte Carlo algorithm which is capable of computing an approximation of the maximum flow in networks with thousands or millions of vertices via subsampling. The algorithm returns both a point estimate of the maximum flow as well as valid Monte Carlo confidence bounds for the true maximum flow. Importantly, its runtime depends only on the number of Monte Carlo samples $B \in \N$ and the size of the subsampled networks, not on the input network. That is, its runtime is $O(B \cdot |\tilde{V}| |\tilde{E}|^2)$, where $B$ is the number of Monte Carlo samples, $\tilde{V}$ is the set of subsampled vertices, and $\tilde{E}$ is the edge set induced by $\tilde{V}$. The runtime can thus be made arbitrarily short or long depending on the desired accuracy encoded in the confidence bounds.

We assess the proposed algorithm with respect to both accuracy and runtime on simulated graphs. Moreover, we demonstrate its capability to compute the maximum flow on large scale networks with hundred thousands of vertices and millions of edges using graphs downloaded from the Brain Networks Data Repository \citep{brainnetworks}.

This article is structured as follows. The proposed algorithm to approximate the maximum flow in large networks is introduced in Section~\ref{sec:methods}, including a specification of the algorithm in pseudocode and its runtime analysis. Experimental results based on simulated networks and networks downloaded from the Brain Networks Data Repository \citep{brainnetworks} are presented in Section~\ref{sec:experiments}. The algorithm concludes with a discussion in Section~\ref{sec:discussion}.

\section{Methods}
\label{sec:methods}
This section introduces our proposed Monte Carlo algorithm based on subsampling to compute the maximum flow of very large graphs. We start with a formal definition of the maximum flow problem (Section~\ref{sec:problem}) before summarizing our main idea (Section~\ref{sec:idea}). We formalize our proposed algorithm in Section~\ref{sec:algorithm}. A runtime analysis can be found in Section~\ref{sec:runtime}.

\subsection{Problem specification}
\label{sec:problem}
We start by formally defining the maximum flow problem. Let $N = (V,E)$ be a capacitated network with source $s \in V$ and sink $t \in V$, where $V$ denotes the set of vertices and $E \subseteq V \times V$ denotes the set of edges. The capacity of an edge is the maximum amount of flow that can pass through it, denoted as $c_{uv}$ for any edge $(u,v) \in E$.

The flow in the network $N$ is a map $f:E \mapsto \R$ satisfying the following two constraints. First, the flow through an edge $(u,v) \in E$, denoted with $f_{uv}$, cannot exceed the capacity of the edge, thus $f_{uv} \leq c_{uv}$ for all $(u,v) \in E$ (capacity constraint). Second, the sum of flows entering any vertex $v$ must equal the sum of flows exiting that vertex, thus $\sum_{u:(v,u) \in E, f_{vu}>0} f_{vu} = \sum_{u:(u,v) \in E, f_{uv}>0} f_{uv}$ for all vertices $v \in V \setminus \{s,t\}$ (flow conservation).

The flow value is then formally defined as the amount of flow passing from the source $s$ to the sink $t$, given by $\sum_{v:(s,v) \in E} f_{sv} = \sum_{u:(u,t) \in E} f_{ut}$. Note that due to the flow conservation property, it suffices to define the flow from the source to the sink in the network as the amount leaving $s$ and entering $t$, respectively (as done in the aforementioned definition). The maximum flow problem asks to find an assignment of flow $f_{sv}$ that maximizes the flow through the network given by $\sum_{v:(s,v) \in E} f_{sv}$.

\subsection{Summary of the idea}
\label{sec:idea}
The idea of our subsampling algorithm is summarized as follows. We are given a network $N=(V,E)$, where $V$ denotes the set of vertices and $E \subseteq V \times V$ denotes the set of edges. We aim to compute the maximum flow from a source $s \in V$ to a sink $t \in V$.

\begin{figure*}
    \begin{minipage}{0.4\textwidth}
        \flushright
        \includegraphics[width=\textwidth]{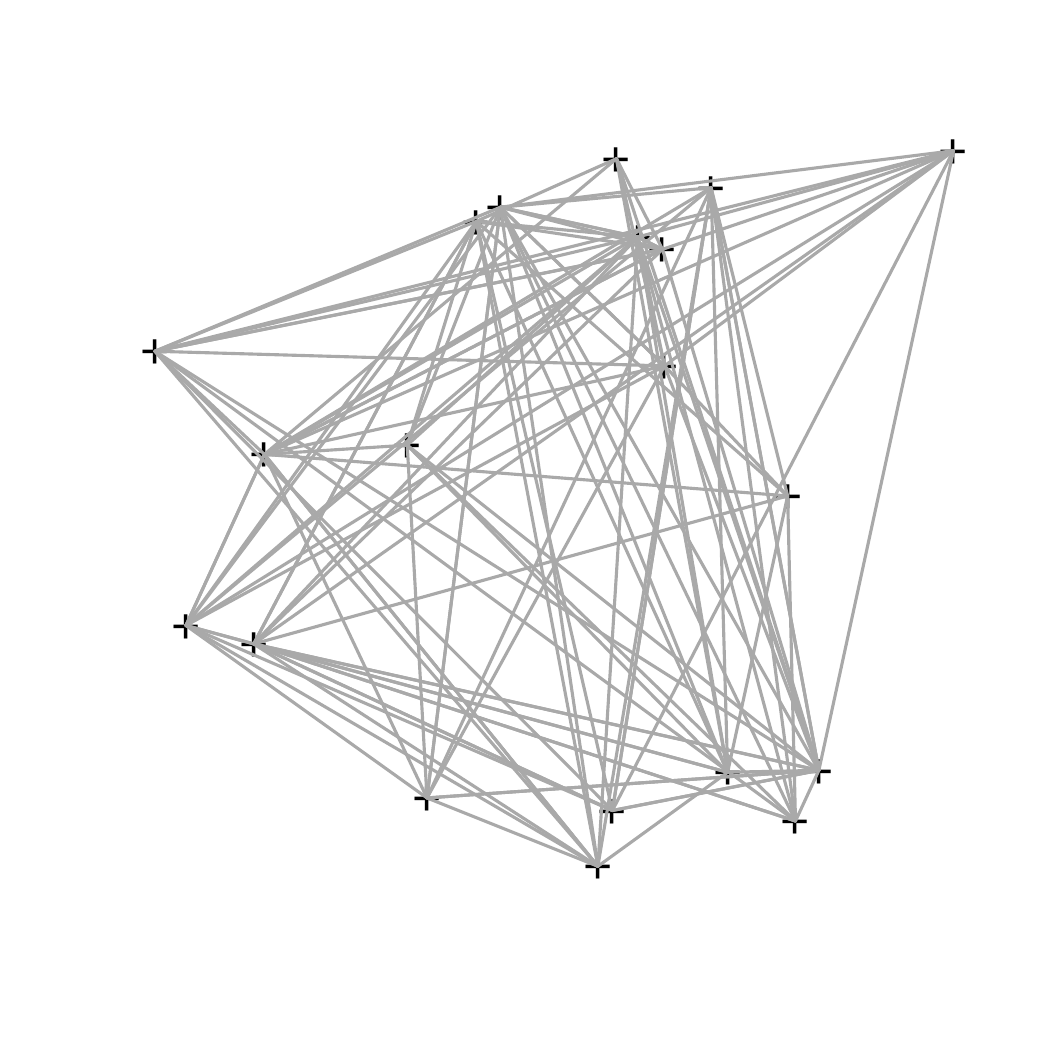}
    \end{minipage}~
    \begin{minipage}{0.15\textwidth}
        \centering
        $\xrightarrow[\text{vertices}]{\mathmakebox[5em]{\text{subsample}}}$
    \end{minipage}~
    \begin{minipage}{0.4\textwidth}
        \flushleft
        \includegraphics[width=\textwidth]{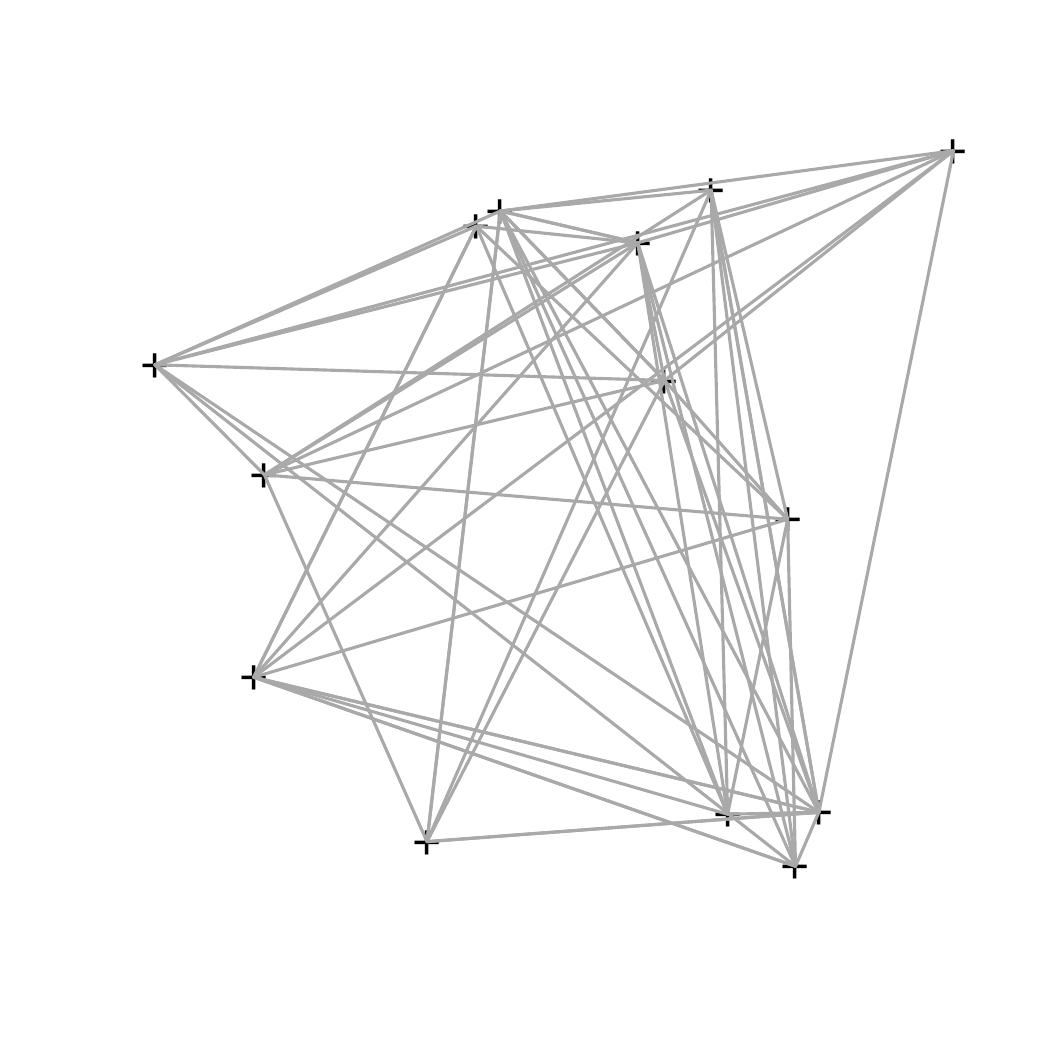}
    \end{minipage}
    \vspace{-4em}

    \begin{minipage}{0.4\textwidth}
        \flushright
        \includegraphics[width=\textwidth]{figure_illustration0.pdf}
    \end{minipage}~
    \begin{minipage}{0.15\textwidth}
        \centering
        $\xrightarrow[\text{edges}]{\mathmakebox[5em]{\text{subsample}}}$
    \end{minipage}~
    \begin{minipage}{0.4\textwidth}
        \flushleft
        \includegraphics[width=\textwidth]{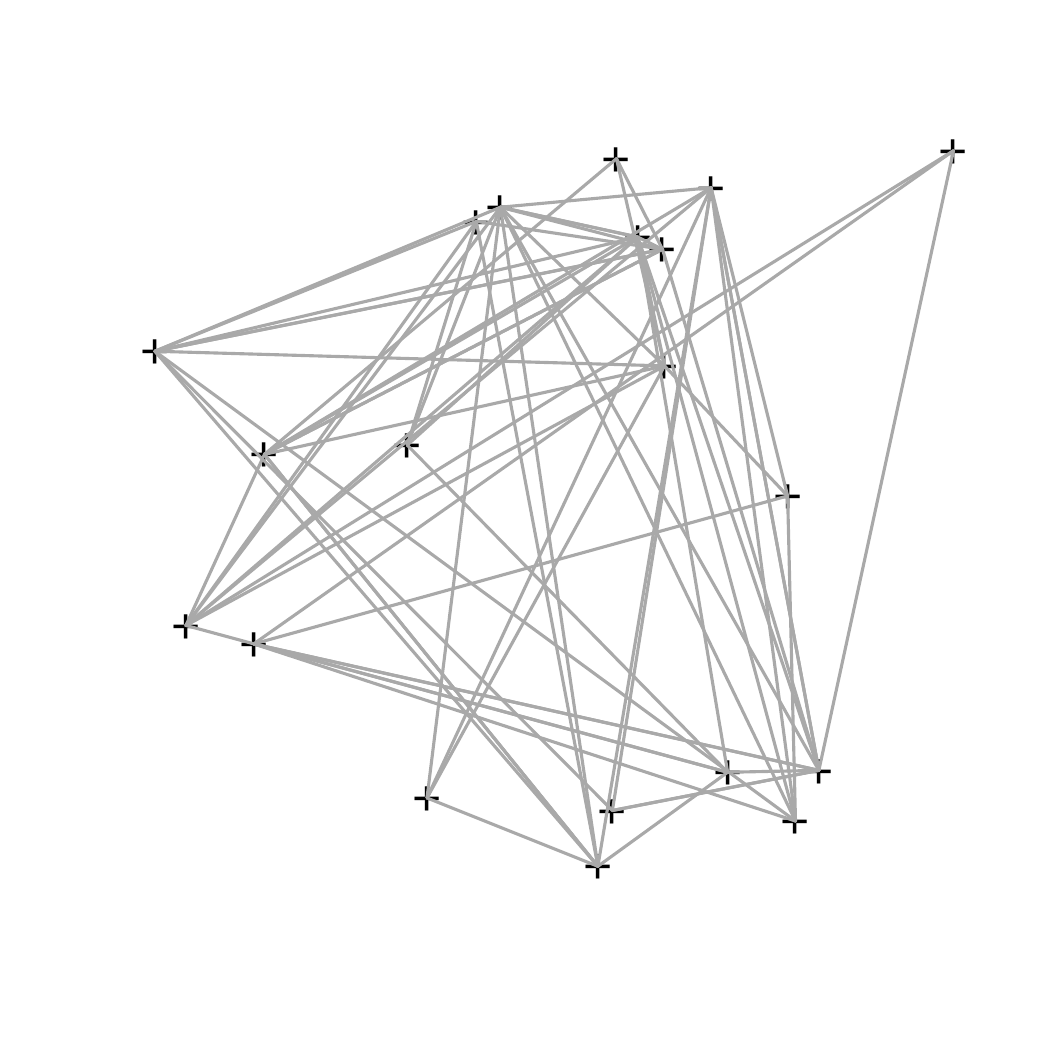}
    \end{minipage}
    \vspace{-2em}
    \caption{Subsamples from a graph $G$ (left) while retaining a fixed proportion of nodes (top) and edges (bottom).}
    \label{fig:illustration}
\end{figure*}

Assume we subsample a proportion $p \in (0,1)$ of the vertices, that is we draw a random subsample $\tilde{V} \subset V$ of size $\lceil p \cdot |V|\rceil$ without replacement while ensuring that $s,t \in \tilde{V}$, where $\lceil \cdot \rceil$ denotes the ceiling operation. Let $\tilde{N}=(\tilde{V},\tilde{E})$ be the subgraph induced by $\tilde{V}$ with implied edge set $\tilde{E} = E \cap (\tilde{V} \times \tilde{V})$. An illustration of a subsampled graph by vertices (or edges) is given in Figure~\ref{fig:illustration}.

When computing a flow in $\tilde{N}$ from $s$ to $t$, we expect the resulting flow value to be on average $p \cdot f$, where $f$ denotes the (full) value of the flow in $N$ from $s$ to $t$. This is due to the fact that $\tilde{N}$ only contains a proportion $p$ of the edges of $N$. Therefore, using the fact that in $\tilde{N}$ we took an unbiased subsample of the edges (and their capacities) of $N$, we would expect the flow $\tilde{f}$ in $\tilde{N}$ to also be a factor $p$ of the original true flow in $\tilde{N}$. We thus project the flow in $\tilde{N}$ onto the original graph $N$ by computing $\tilde{f}/p$, which is our estimate for the true flow $f$ in $N$. This procedure can be repeated an arbitrary number of times $B \in \N$, the number of Monte Carlo subsamples we take from $N$. At the end, we obtain $B$ Monte Carlo estimates of the maximum flow. We can summarize those in a straightforward way by computing their mean and standard deviation (for instance, to yield a normal confidence interval).

\subsection{The subsampling algorithm}
\label{sec:algorithm}
The idea outlined in Section~\ref{sec:idea} leads to the method presented in Algorithm~\ref{algo:subsampling}. Algorithm~\ref{algo:subsampling} receives a capacitated and directed network $N=(V,E)$ with vertex set $V$ and edge set $E \subseteq V \times V$, as well as the source and sink vertices $s,t \in V$, the subsampling proportion $p \in (0,1)$, and the number of Monte Carlo samples $B \in \N$.

\begin{algorithm}[t]
    \caption{Maximum flow via subsampling}
    \label{algo:subsampling}
    \begin{algorithmic}[1]
    \Require network $N=(V,E)$, source $s \in V$, sink $t \in V$, subsample proportion $p \in (0,1)$, number of Monte Carlo replicates $B \in \N$
    \State Initialize vector $\phi \in \R^B$
    \For{$i \in \{1,\ldots,B\}$}
        \State $\sigma \leftarrow \lceil p \cdot |V|\rceil$
        \State $\tilde{V} \leftarrow$ random subsample of size $\sigma-2$ from $V \setminus \{s,t\}$ without replacement
        \State $\tilde{V} \leftarrow \tilde{V} \cup \{s,t\}$
        \State $\tilde{N} \leftarrow$ subgraph of $N$ induced by vertex set $\tilde{V}$ and implied edge set $\tilde{E} = E \cap (\tilde{V} \times \tilde{V})$
        \State $\tilde{f} \leftarrow$ flow from $s$ to $t$ in $\tilde{N}$
        \State $\phi_i \leftarrow \tilde{f}/p$
    \EndFor
    \State \Return $\textnormal{mean}(\phi)$ and $\textnormal{sd}(\phi)$
    \end{algorithmic}
\end{algorithm}

The algorithm subsamples $\tilde{V}$ of size $\sigma = \lceil p \cdot |V|\rceil$ from $V$ while making sure that $s,t \in V$ are always contained in $\tilde{V}$. The maximum flow $\tilde{f}$ is computed on the subgraph $\tilde{N}$ induced by $\tilde{V}$ and $\tilde{E} = E \cap (\tilde{V} \times \tilde{V})$. Since $\tilde{f}$ is expected to approximate $p \cdot f$, we obtain an estimate of the full flow as $f \approx \tilde{f}/p$ and store it in a vector $\phi \in \R^B$. This procedure is repeated $B$ times and the resulting mean and standard deviation are returned.

\subsection{Runtime analysis}
\label{sec:runtime}
The runtime of Algorithm~\ref{algo:subsampling} is determined by three factors, the size of the subsampled vertex sets $\tilde{V}$, the algorithm chosen to compute the maximum flow on the subsampled network, and the number of Monte Carlo replicates $B$.

We assume that the flow in $\tilde{N}$ is computed with the Edmonds-Karp algorithm \citep{Edmonds1972} in time $O(|\tilde{V}||\tilde{E}|^2)$. For $B$ Monte Carlo replicates, this amounts to a total runtime of $O(B \cdot |\tilde{V}| |\tilde{E}|^2)$. Therefore, the runtime of Algorithm~\ref{algo:subsampling} is independent of the actual vertex and edge sets $V$ and $E$. That is, by appropriately subsampling them it is possible to get an estimate of the maximum flow for any desired computational effort. However, this usually comes at the cost of a reduced accuracy of the maximum flow value.

\section{Experimental results}
\label{sec:experiments}
This section presents our experimental results. We start with a brief overview of the experimental setting in Section~\ref{sec:setting}, followed by an assessment of Algorithm~\ref{algo:subsampling} with respect to its accuracy in Section~\ref{sec:behavior}. Moreover, we investigate the relative error of Algorithm~\ref{algo:subsampling} (Section~\ref{sec:relerror}) and its scaling behavior for large networks (Section~\ref{sec:scaling}). We conclude with experimental results on large real-world networks downloaded from the Brain Networks Data Repository (Section~\ref{sec:large_networks}).

\subsection{Experimental setting}
\label{sec:setting}
Throughout the experiments, we generate random graphs using the Erd{\H o}s-R{\'e}nyi model \citep{Erdos1959}. Erd{\H o}s-R{\'e}nyi random graphs have two parameters, the number of vertices $n \in \N$ and the edge probability $\pi \in [0,1]$, meaning each edge is independently drawn with probability $\pi$. The size of the vertex set is given separately for each experiment. We assign weights to all edges by drawing them uniformly at random from the integer set $\{1,\ldots,10\}$. In each generated random graph, we always select two vertices at random and assign them to be the source and the sink, respectively.

Whenever we compute a maximum flow, we employ the classic Edmonds-Karp algorithm \citep{Edmonds1972}. The Edmonds-Karp algorithm is run using the function \textit{maximum\_flow\_value} of the \textit{networkx} library in Python.

All code was implemented on an Apple M2 Pro Chip running macOS Ventura 13.4, using Python 3.11.5 compiled with Clang 14.0.6.

\subsection{Behavior of the subsampling algorithm}
\label{sec:behavior}
We investigate the behavior of Algorithm~\ref{algo:subsampling} on four Erd{\H o}s-R{\'e}nyi random graphs with $|V|=1000$ vertices and edge probabilities $\pi \in \{0.2,0.4,0.6,0.8\}$. We measure the estimated maximum flow value as a function of the subsampling proportion $p$. All confidence bands are based on $50$ replicates. Additionally, we also compute the exact maximum flow with the help of the Edmonds-Karp algorithm. Since the maximum flow differs among the networks created with different edge probabilities $\pi \in \{0.2,0.4,0.6,0.8\}$, we calculate the percentage of the exact flow recovered with Algorithm~\ref{algo:subsampling}.

\begin{figure*}
    \centering
    \includegraphics[width=0.49\linewidth]{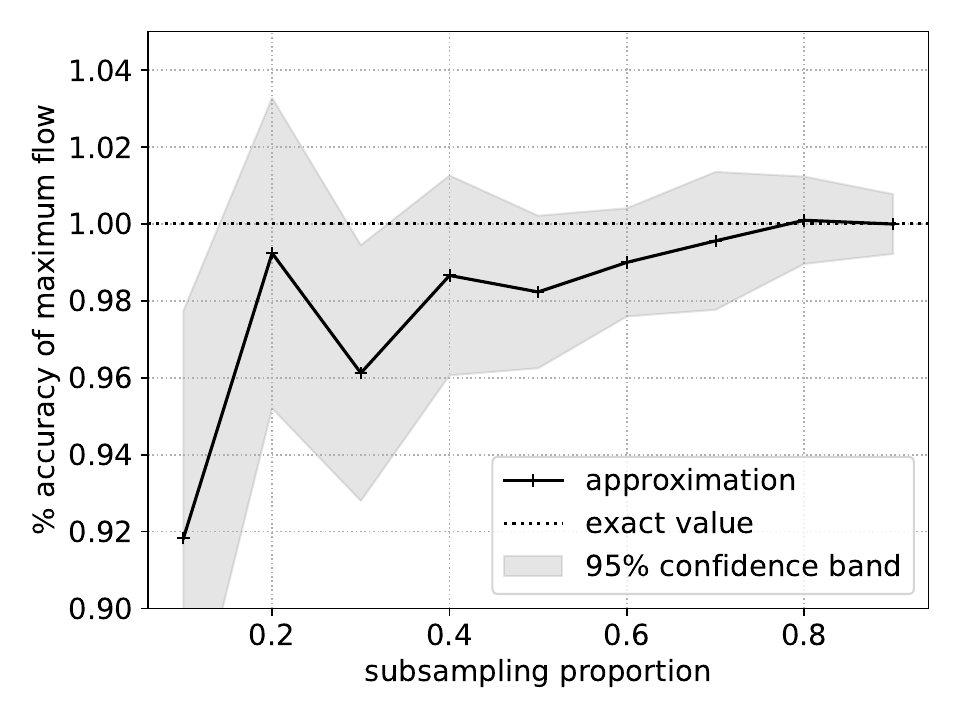}~
    \includegraphics[width=0.49\linewidth]{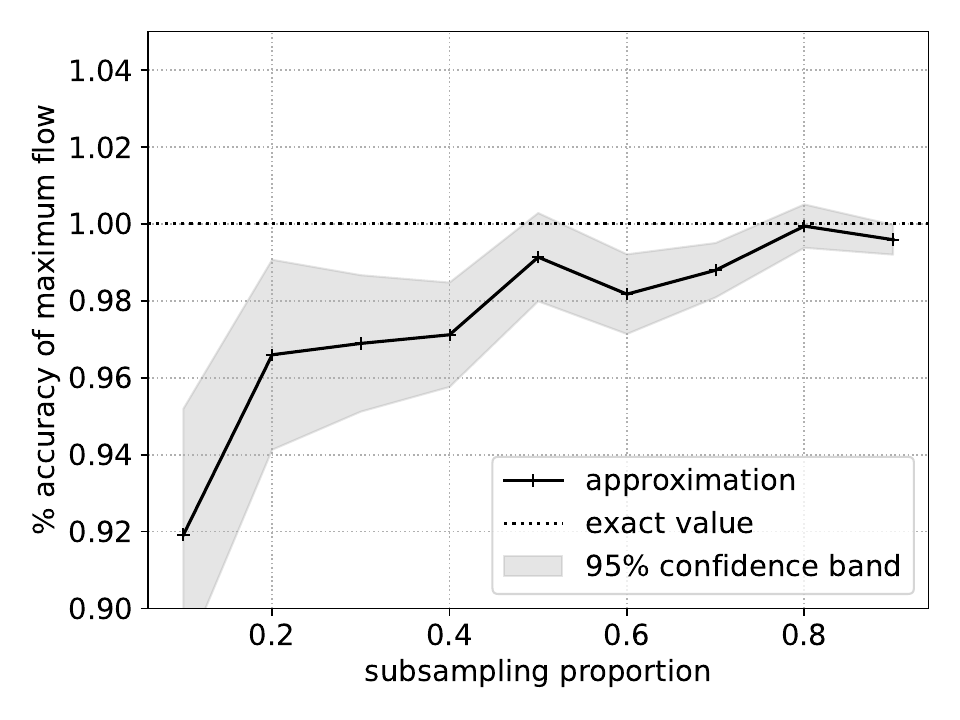}\\
    \includegraphics[width=0.49\linewidth]{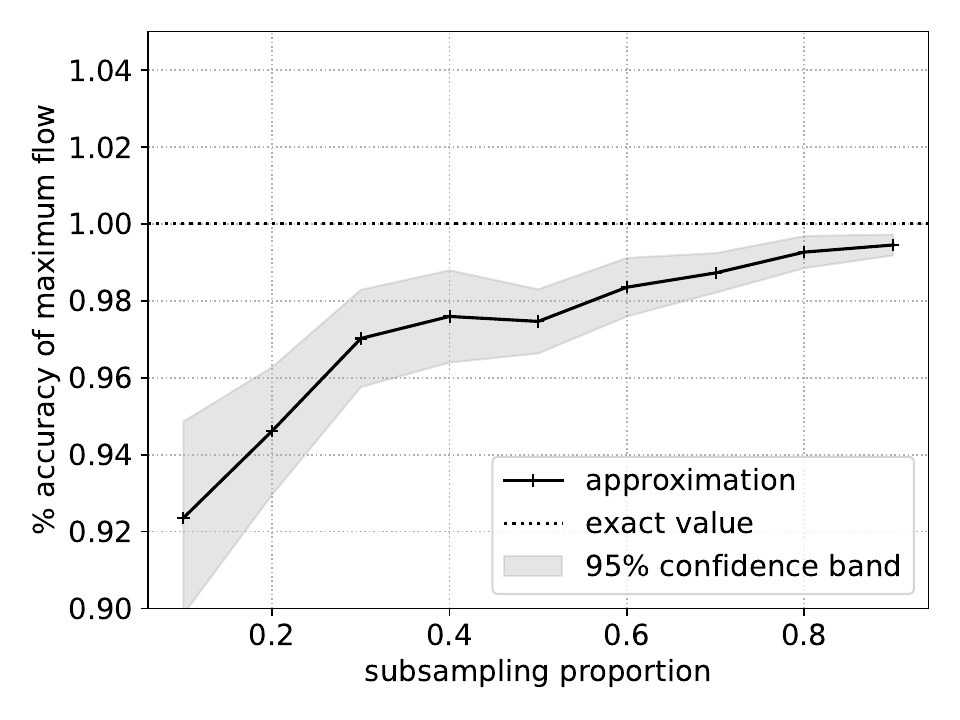}~
    \includegraphics[width=0.49\linewidth]{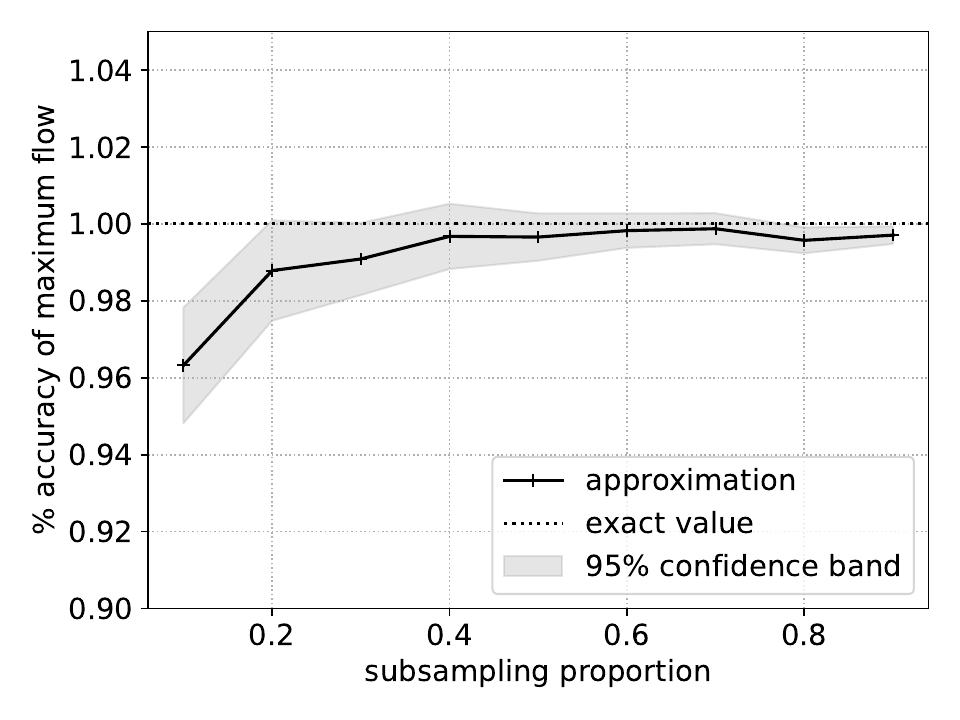}\\
    \caption{Accuracy of the maximum flow estimates (percentage of exact flow) of Algorithm~\ref{algo:subsampling} as a function of the subsampling proportion $p$. Four Erd{\H o}s-R{\'e}nyi random graphs with $|V|=1000$ vertices and edge probabilities $\pi=0.2$ (top left), $\pi=0.4$ (top right), $\pi=0.6$ (bottom left), and $\pi=0.8$ (bottom right). Confidence bands depicted in gray.}
    \label{fig:example}
\end{figure*}

Figure~\ref{fig:example} shows the results of this experiment. Several observations are noteworthy. First, we observe across all four plots that the accuracy of the estimated flow increases with the subsampling proportion. This behavior is expected, as for $p \rightarrow 1$, Algorithm~\ref{algo:subsampling} transforms into a repeated application of Edmonds-Karp on the original graph. Overall, we observe that the maximum flow estimates returned by Algorithm~\ref{algo:subsampling} are in the neighborhood of $96\%$ accuracy for subsampling proportions of $p>0.2$.

Second, we observe that the maximum flow estimates obtained with Algorithm~\ref{algo:subsampling} are more volatile for smaller edge probabilities $\pi$ (as visible from the larger confidence bands). This is expected, because subsampling networks with fewer edges more likely results in networks with different characteristics than the original network. In turn, the maximum flow estimates are considerably less volatile for dense networks.

Third, the accuracy of our algorithm is higher in general for dense networks than for sparse ones. This means that Algorithm~\ref{algo:subsampling} approximates the maximum flow for lower subsampling proportions more accurately for dense than for sparse networks. This can be explained by the fact that dense networks allow for a sufficient maximum flow through them that is representative of the exact flow even when some vertices and edges are removed.

\subsection{Relative error}
\label{sec:relerror}
Similarly to Section~\ref{sec:behavior}, we aim to investigate the relative error of Algorithm~\ref{algo:subsampling} as a function of the subsampling proportion $p$. For this we generate Erd{\H o}s-R{\'e}nyi random graphs with $|V|=1000$ vertices and edge probabilities (densities) $\pi \in \{0.1,0.3,0.5,0.7,0.9\}$.

\begin{figure*}
    \centering
    \includegraphics[width=0.5\linewidth]{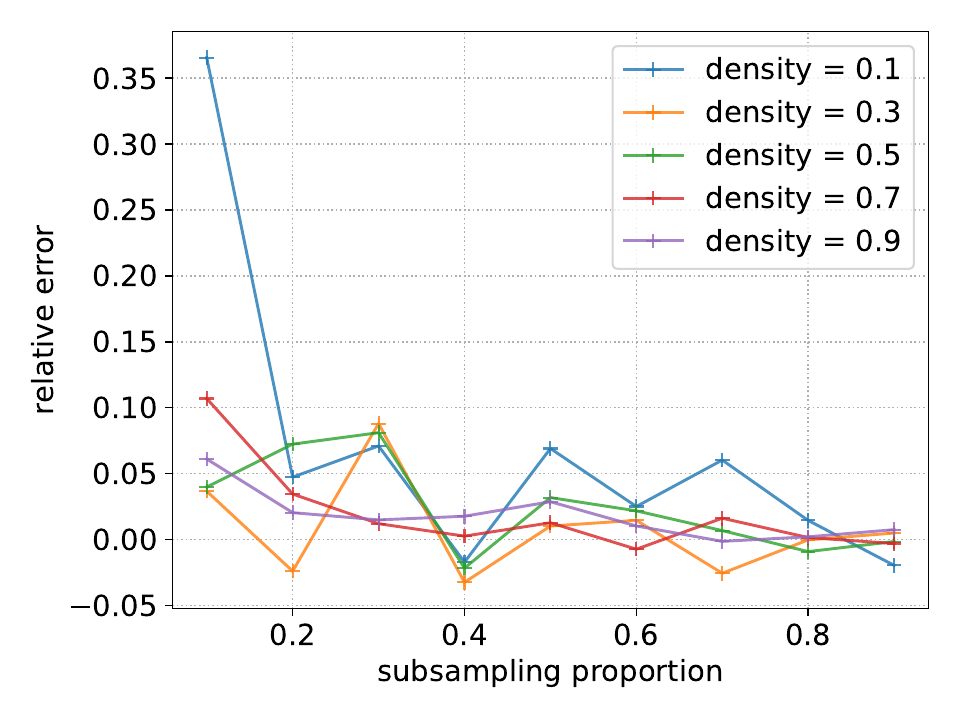}
    \caption{Relative error of the maximum flow estimates of Algorithm~\ref{algo:subsampling} as a function of the subsampling proportion $p$. Erd{\H o}s-R{\'e}nyi random graphs with $|V|=1000$ vertices and edge probabilities (densities) $\pi \in \{0.1,0.3,0.5,0.7,0.9\}$.}
    \label{fig:relerror}
\end{figure*}

Figure~\ref{fig:relerror} shows the relative error as a function of the subsampling proportion. We observe that across all densities, given a subsampling proportion $p>0.2$, the relative error of the maximum flow estimate returned by Algorithm~\ref{algo:subsampling} quickly drops to less than $5\%$ (both to the above and below) of the exact maximum flow value.

\subsection{Scaling for large networks}
\label{sec:scaling}
We are interested in the runtime scaling for both the Edmonds-Karp algorithm and Algorithm~\ref{algo:subsampling} as a function of the number of vertices $|V|$. To this end, we generate Erd{\H o}s-R{\'e}nyi random graphs with a varying number of vertices $|V| \in \{1000,5000,10000,50000,100000\}$ and fixed edge probability $\pi=0.001$. We apply Algorithm~\ref{algo:subsampling} with a subsampling proportion tied to the graph size, given as $p=\min(0.5,10000/|V|)$. Computations were stopped after $60$ seconds. We record the obtained maximum flow value and the runtime.

\begin{figure*}
    \centering
    \includegraphics[width=0.49\linewidth]{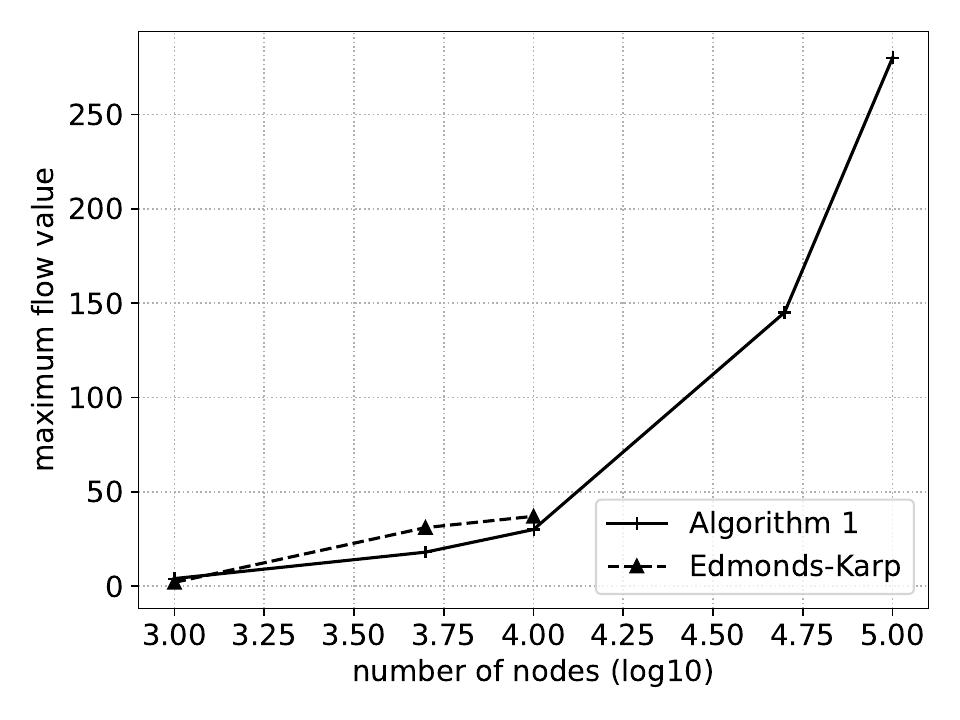}~
    \includegraphics[width=0.49\linewidth]{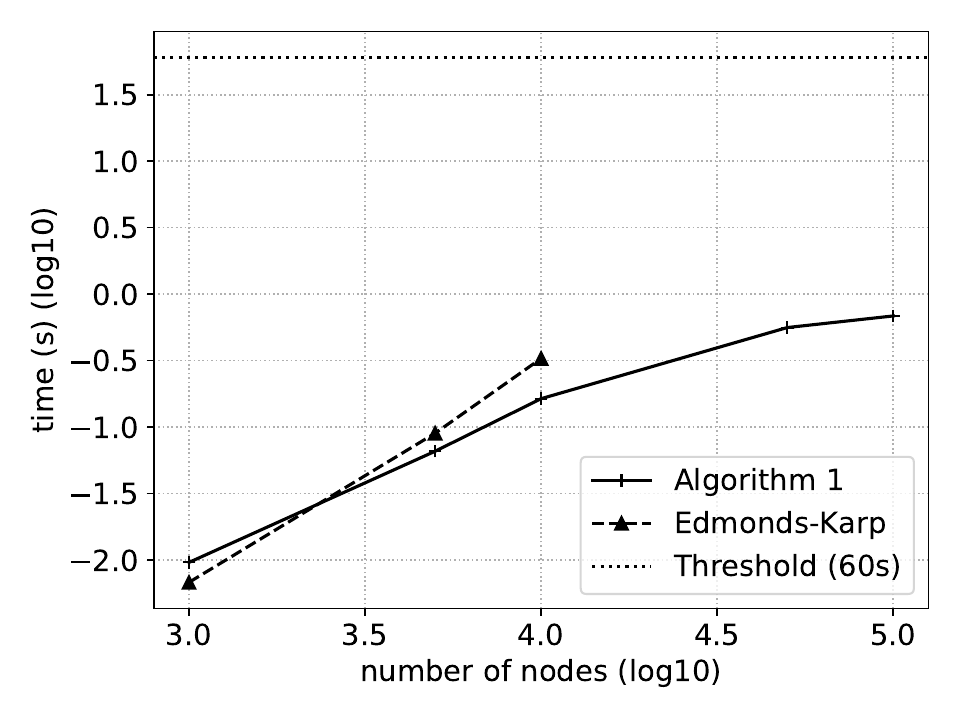}
    \caption{Maximum flow value (left) and runtime in seconds (right) as a function of the number of vertices $|V|$. Erd{\H o}s-R{\'e}nyi random graphs with $|V|$ given on the x-axis and fixed edge probability $\pi=0.001$. The subsampling proportion for Algorithm~\ref{algo:subsampling} was chosen as $p=\min(0.5,10000/|V|)$. Computations were stopped after $60$ seconds. Log scale on the x-axis (left) and on both axes (right).}
    \label{fig:scaling}
\end{figure*}

Figure~\ref{fig:scaling} displays both the maximum flow value and the runtime as a function of the network size. First, we observe that the Edmonds-Karp algorithm was not able to compute a maximum flow for the two largest network sizes, while Algorithm~\ref{algo:subsampling} returns approximate values that are close to the ones of Edmonds-Karp (where available) and reasonable otherwise. Second, Algorithm~\ref{algo:subsampling} has a more favorable runtime scaling, with slope $0.94$ in the log-log plot compared to Edmonds-Karp (slope $1.67$). We note that the experimentally measured slope for the Edmonds-Karp algorithm lies below the theoretical runtime dependence, a fact we attribute to its optimized implementation in the \textit{networkx} library.

\subsection{Application to large real-world networks}
\label{sec:large_networks}
Finally, we apply both the Edmonds-Karp algorithm and Algorithm~\ref{algo:subsampling} to six real-world brain networks published in the Brain Networks Data Repository \cite{brainnetworks}. The online references to those networks can be found in Appendix~\ref{sec:availability}.

\begin{table*}
    \centering
    \begin{tabular}{l|rr|rr|rr}
        \hline
        Dataset name & Nodes & Edges & Algorithm~\ref{algo:subsampling} & EK & Algorithm~\ref{algo:subsampling} & EK\\
        & & & Time & Time & Value & Value \\
        \hline
        BNU-1-0025890 & 177584 & 15669037 & \textbf{75.17} & 136.36 & 85.24 & 92\\
        BNU-1-0025890 & 177584 & 15669037 & \textbf{66.79} & 176.70 & 14.21 & 20\\
        Jung2015-M87127677 & 223969 & 38543660 & \textbf{138.43} & 502.00 & 201.57 & 224\\
        Jung2015-M87126947 & 268076 & 46092588 & \textbf{95.30} & 263.63 & 37.53 & 32\\
        Jung2015-M87128519 & 395665 & 59883534 & \textbf{97.13} & N/A & 51.44 & N/A\\
        Jung2015-M87118347 & 428842 & 79114771 & \textbf{98.14} & 434.31 & 8.58 & 5\\
        \hline
    \end{tabular}
    \caption{Performance of the Edmonds-Karp (EK) algorithm and Algorithm~\ref{algo:subsampling} on graphs downloaded from the Brain Network Data Repository. The table shows the network name, its nodes and edges, as well as the runtime and maximum flow value measured with both algorithms. A value of N/A indicates that Edmonds-Karp was unable to compute the maximum flow within the time limit of 300 seconds.}
    \label{tab:large_networks}
\end{table*}

Table~\ref{tab:large_networks} displays the network name as well as its number of vertices and edges. It also displays the runtime and maximum flow value found with both algorithms, where N/A indicates that Edmonds-Karp was unable to compute the maximum flow within the time limit of 300 seconds. Comparing the values of the maximum flow in the last two columns, we observe that Algorithm~\ref{algo:subsampling} is able to return approximate maximum flows for these networks that are reasonably close to the exact values. At the same time, Algorithm~\ref{algo:subsampling} achieves this with roughly a two-fold to five-fold speedup compared to Edmonds-Karp.

\section{Discussion}
\label{sec:discussion}
In this contribution, we introduce a simple approximation algorithm for the maximum flow in large networks. The algorithm is based on the idea that random subsamples have a similar connectivity structure than the original network, and thus that the maximum flow computed in a subsample can be scaled back to the flow in the full network. Repeatedly computing the maximum flow on several subsampled networks allows one to compute an estimate and a confidence interval of the maximum flow. We assess the behavior of our proposed algorithm in a simulation study.

Importantly, the runtime of our proposed algorithm technically does not depend on the size of the input network, since the maximum flow in any graph can in principle be approximated using subsamples of fixed size. In practice, however, we would make the size of the subsampled networks depended on the input. In our simulation study, we show that subsampling proportions as low as $p=0.2$ give results with an error margin of $5\%$. To be precise, the runtime of our method is $O(B \cdot |\tilde{V}| |\tilde{E}|^2)$, where $B$ is the number of Monte Carlo samples, $\tilde{V}$ is the set of subsampled vertices, and $\tilde{E} = E \cap (\tilde{V} \times \tilde{V})$ is the edge set induced by $\tilde{V}$. The runtime can thus be made arbitrarily short or long depending on the desired accuracy encoded in the confidence bounds.

\subsection*{Acknowledgements}
The authors gratefully acknowledge the contributors of the Brain Networks Data Repository.

\appendix
\section{Availability of datasets}
\label{sec:availability}
The datasets analyzed in this study are available from the Brain Networks Data Repository, see \url{https://networkrepository.com/}. Their URL identifiers are given in Table~\ref{tab:links}.

\begin{table*}
    \small
    \begin{tabular}{ll}
    \hline
    BNU-1-0025890 & \url{https://networkrepository.com/bn-human-BNU-1-0025890-session-1.php}\\
    BNU-1-0025890 & \url{https://networkrepository.com/bn-human-BNU-1-0025890-session-2.php}\\
    Jung2015-M87127677 & \url{https://networkrepository.com/bn-human-Jung2015-M87127677.php}\\
    Jung2015-M87126947 & \url{https://networkrepository.com/bn-human-Jung2015-M87126947.php}\\
    Jung2015-M87128519 & \url{https://networkrepository.com/bn-human-Jung2015-M87128519.php}\\
    Jung2015-M87118347 & \url{ https://networkrepository.com/bn-human-Jung2015-M87118347.php}\\
    \hline
    \end{tabular}
    \caption{List of URL identifiers for all networks downloaded from the Brain Networks Data Repository \cite{brainnetworks}.}
    \label{tab:links}
\end{table*}

\bibliographystyle{apalike}
\bibliography{main}

\end{document}